\documentclass[11pt,twoside]{article}
\usepackage{asp2004}
\usepackage{psfig}
\usepackage{epsf}
\usepackage{graphics}
\usepackage{lscape}
\begin{document}
\title{The enigmatic B[e] star Hen 2-90 -- an interacting binary?}
\author{M. Kraus}
\affil{Astronomical Institute, Utrecht University, Princetonplein 5, 3584 CC Utrecht, The Netherlands}
\author{M. Borges Fernandes}
\affil{Observat\'orio do Valongo, Universidade Federal do Rio de Janeiro, 
Ladeira do Pedro Ant\^onio 43, 20080-090, Sa\'ude, Rio de Janeiro, Brasil}
\author{F.X. de Ara\'{u}jo}
\affil{Observat\'{o}rio Nacional-MCT, Rua General Jos\'{e} Cristino 77, 20921-400 S\~{a}o Cristov\~{a}o, Rio de Janeiro, Brasil}

\begin{abstract}
We present the enigmatic galactic unclassified B[e] star Hen 2-90. Its 
optical spectrum is discussed and analysed in terms of the forbidden 
emission lines coming from a non-spherical wind. The evolutionary phase
of Hen 2-90 is discussed. We raise the question whether an interacting binary 
nature seems to be a possible solution to account for all observed 
characteristics.  
\end{abstract}
\thispagestyle{plain}

\section{Introduction}

Stars with the B[e] phenomenon have besides their spectral type several
characteristics in common: strong Balmer emission lines, many permitted and 
forbidden emission lines of low ionized metals, e.g. [Fe{\sc ii}] and [O{\sc 
i}], and a strong near- or mid-infrared excess due to hot circumstellar dust.

The group of stars showing the B[e] phenomenon is heterogeneous and has been 
divided by \citet{l1998} into subgroups according to their 
evolutionary phase. These subgroups contain supergiants, Herbig stars, 
symbiotic objects and compact planetary nebulae. The biggest group, however, 
are the unclassified B[e] stars whose evolutionary phase is not or not 
unambiguously known.

Here, we present the enigmatic object Hen 2-90, a galactic
unclassified B[e] star. It has first been classified as a 
compact planetary nebula, e.g. by \citet{h1967}; \citet{s1992}; \citet{c1993};
\citet{l1998}. Later on, a bipolar jet-like structure was resolved with HST 
\citep{sn2000} with several perfectly aligned knots extending up to 
$\sim 10\arcsec$~on both sides of the star. Such a structure reminds more of 
a young stellar object or a symbiotic object than of a compact planetary 
nebula. Further observations with HST \citep{s2002} revealed even a bipolar 
high-ionized wind, a low-ionized wind at intermediate latitudes as well as a 
high-density circumstellar disk.

\section{The nature of Hen 2-90}

Due to the different classifications found in the literature we wanted to
test their reliability. In this section we therefore discuss the different
possible classifications in terms of characteristics of Hen 2-90 found from
already existing observations and found from our own set of observations.

\subsection{Non-spherical mass loss in a compact planetary nebula?}  

The circumstellar material on the HST image of Hen 2-90 \citep{s2002} shows a
latitude dependence of the ionization structure turning from high-ionization in polar directions to low-ionization in equatorial directions. Such a
behaviour might be explained with a latitude dependent mass flux and surface
temperature distribution as a result of a rapidly rotating underlying star
\citep{k2004}.
\begin{table}[!h]
\caption{Observed and modelled forbidden line luminosities for a stellar 
distance of 2\,kpc. Column 4 gives the ratio of observed over modelled line 
luminosity, Column 5 the wind part(s) in which 
the line is produced: polar (p), intermediate (i) or disk wind 
(d).}\label{linelum}
\smallskip
\begin{center}
{\small
\begin{tabular}{lccccc}
\tableline
\noalign{\smallskip}
Ion & $\lambda$ (\AA)  &   $L_{\lambda}^{\rm obs}$  &  $L_{\lambda}^{\rm
model}$ & ratio & region  \\
\noalign{\smallskip}
\tableline
\noalign{\smallskip}
O{\sc iii}  & 4959 & $3.58\times 10^{34}$ & $3.91\times 10^{34}$ & 0.92 & p,i \\
O{\sc iii}  & 5007 & $1.13\times 10^{35}$ & $1.13\times 10^{35}$ & 1.00 & p,i \\
O{\sc iii}  & 4363 & $9.71\times 10^{33}$ & $2.90\times 10^{34}$ & 0.33 & p,i \\
O{\sc ii}   & 7319 & $7.23\times 10^{33}$ & $7.70\times 10^{33}$ & 0.94 & i \\
O{\sc ii}   & 7330 & $6.18\times 10^{33}$ & $6.17\times 10^{33}$ & 1.00 & i \\
O{\sc i}    & 6300 & $4.98\times 10^{32}$ & $4.86\times 10^{32}$ & 1.02 & d \\
O{\sc i}    & 6364 & $1.49\times 10^{32}$ & $1.60\times 10^{32}$ & 0.93 & d \\
O{\sc i}    & 5577 & $1.12\times 10^{32}$ & $2.33\times 10^{32}$ & 0.48 & d \\
S{\sc iii}  & 6312 & $2.95\times 10^{33}$ & $2.98\times 10^{33}$ & 0.99 & i \\
S{\sc ii}   & 6731 & $2.13\times 10^{32}$ & $2.20\times 10^{32}$ & 0.97 & i,d \\
S{\sc ii}   & 6716 & $9.05\times 10^{31}$ & $1.05\times 10^{32}$ & 0.86 & i,d \\
S{\sc ii}   & 4076 & $3.43\times 10^{32}$ & $1.18\times 10^{32}$ & 2.90 & i,d \\
S{\sc ii}   & 4068 & $1.15\times 10^{33}$ & $4.57\times 10^{32}$ & 2.52 & i,d \\
N{\sc ii}   & 5755 & $4.56\times 10^{33}$ & $4.57\times 10^{33}$ & 1.00 & i \\
N{\sc ii}   & 6548 & $4.84\times 10^{33}$ & $6.15\times 10^{33}$ & 0.79 & i \\
N{\sc ii}   & 6584 & $2.08\times 10^{34}$ & $1.81\times 10^{34}$ & 1.15 & i \\
Cl{\sc iii} & 5538 & $1.74\times 10^{32}$ & $1.74\times 10^{32}$ & 1.00 & p,i \\
Cl{\sc iii} & 5517 & $1.10\times 10^{32}$ & $4.23\times 10^{31}$ & 2.60 & p,i \\
Cl{\sc ii}  & 6153 & $3.88\times 10^{31}$ & $3.82\times 10^{31}$ & 1.01 & i \\
Ar{\sc iii} & 7136 & $7.37\times 10^{33}$ & $6.88\times 10^{33}$ & 1.07 & p,i \\
Ar{\sc iii} & 7753 & $1.49\times 10^{33}$ & $1.68\times 10^{33}$ & 0.89 & p,i \\
Ar{\sc iii} & 5193 & $1.16\times 10^{32}$ & $6.44\times 10^{32}$ & 0.18 & p,i \\
\noalign{\smallskip}
\tableline
\end{tabular}
}
\end{center}
\end{table}
We took optical spectra\footnote{Based on observations with the
1.52m telescope at the European Southern Observatory (La Silla, Chile), under
the agreement with the Observat\'{o}rio Nacional-MCT (Brasil)} centered on the
star with a slit width that just covers the non-spherical ionization
structure visible in the HST image of \citet{s2002}.

The spectrum shows strong [O{\sc i}] emission which is an indication for
the presence of a huge amount of hydrogen neutral material close to the star
because H and O have the same ionization potential. According to the HST image,
this neutral material can only be located in the equatorial disk.
As \citet{k2003} showed in the case of B[e] supergiants, equatorial disks
around hot stars can indeed be neutral even close to the hot stellar surface,
simply due to the high equatorial mass fluxes that result in effective
shielding of the disk material from the ionizing stellar continuum photons.

Besides the O lines, the optical spectrum of Hen 2-90 also contains much more 
forbidden emission lines that can be related with the different wind regions 
(see Table \ref{linelum}). We performed a detailed analysis of these 
forbidden emission lines and found mass fluxes that increase from the pole to 
the equator by about a factor of 8 while the surface 
temperature drops by about a factor 3 \citep[for more details see][]{k2004}. 
These facts seem to be consistent with a rotating central star model. However, 
to fit the observed line luminosities, we had to adopt the elemental 
abundances. We found that S, Ar, Cl and He have about solar abundances, while 
O seems to be depleted (0.3 solar). We did not identify any C line which 
speaks in favour of C being depleted, too. These abundances are in agreement 
with Hen 2-90 being an evolved single star. The puzzling thing, however, is the 
fact, that also N seems to be depleted (0.5 solar) which cannot be explained by 
standard stellar evolution. We conclude that our simple picture 
of a rotating single star undergoing non-spherical mass loss cannot be the full
story.

\subsection{The jet and knots of Hen 2-90}

The strange abundances found from our modeling and the fact that Hen 2-90
has a jet-like structure with perfectly aligned knots which are ejected
regularly might help coming closer to the real nature of this puzzling and in a 
sense unique object. In Table\,\ref{knots} we listed all objects known to 
possess jets and knots and the identification of their source of jet and knot 
formation.

\begin{table}[!b]
\caption{Objects showing jets and knots \citep[after][]{l1999}.}\label{knots}
\smallskip
\begin{center}
{\small
\begin{tabular}{ll}
\tableline
\noalign{\smallskip}
Object & Physical system \\
\noalign{\smallskip}
\tableline
\noalign{\smallskip}
Young stellar objects & {\it Accreting} young star \\
Massive X-ray binaries & {\it Accreting} neutron star or black hole \\
Black hole X-ray transients & {\it Accreting} black hole \\
Low mass X-ray binaries & {\it Accreting} neutron star \\
Supersoft X-ray sources & {\it Accreting} white dwarf \\
Symbiotic stars & {\it Accreting} white dwarf \\
\noalign{\smallskip}
\tableline
\noalign{\smallskip}
Hen 2-90 & {\it Accreting} ????? \\
\noalign{\smallskip}
\tableline
\end{tabular}
}
\end{center}
\end{table}

From this table we can draw two major conclusions: (1) except of the 
young stellar objects, all jet systems are binaries, and (2)
the jet and knot appearence is always linked to some kind of accretion.
In the following sections, we therefore want to discuss 
whether the characteristics of Hen 2-90 fit into one of the classes of objects
listed in Table\,\ref{knots}.

\subsection{Hen 2-90 -- a young stellar object?}

The perfectly aligned knots on both sides of the star remind of a Herbig
object, i.e. a young stellar object (YSO). There are, however, a few 
points that speak against the identification of Hen 2-90 as a YSO:
\begin{itemize}
\item Hen 2-90 is not located in a star forming region,
\item its IRAS colors are much hotter than those of a YSO, separating Hen 2-90
in a color-color diagram from the regions covered by YSOs, OH/IR stars, 
H{\sc ii} regions, and ultracompact H{\sc ii} regions, and
\item the depletion in C and O found from our modeling speak more in
favour of an evolved object rather than a YSO.
\end{itemize}
To us it seems therefore clear that Hen 2-90 cannot be a YSO. If it belongs
to one of the categories defined in Table\,\ref{knots}, Hen 2-90 must be 
a binary.

\subsection{An X-ray binary?}

Most of the binaries in Table\,\ref{knots} are X--ray sources. \citet{c2003} 
studied the X--ray emission from planetary nebulae and wind blown bubbles and 
superbubbles. Hen 2-90 was on their list of Chandra observations, but could not 
be detected in the 0.1--10\,keV range. As far as we know, this was the only 
investigation of Hen 2-90 in X--rays. Observations e.g. in the hard X-ray
band are certainly needed to quantify whether Hen 2-90 is an X--ray source.

\subsection{A symbiotic object?}

A symbiotic object is a binary consisting of a hot component, normally a white
dwarf, and a cool component, normally a giant star. The optical spectrum 
of a symbiotic is a combination of these two individual spectra and therefore 
contains characteristics of both, the hot and the cool component.
The major characteristics of symbiotic objects are:
\begin{itemize}
\item strong He{\sc ii} emission from the hot component
\item TiO absorption bands arising in the atmosphere of the cool giant
\end{itemize}
Our optical spectrum shows indication neither for He{\sc ii} emission, nor for 
TiO absorption bands. These features might be hidden within the circumstellar
disk-like material, but the clear absence of any emission line coming from
ions with ionization potential larger than 40\,eV speaks more in favour of
that these ions do not exist, which means that the ionizing source in Hen 2-90
cannot be too hot, which is consistent with the fact that the effective 
temperature has been found to be of order 50\,000\,K \citep{k1991}. We 
therefore exclude Hen 2-90 being a symbiotic object.

\section{Conclusions}

We presented the unclassified B[e] star Hen 2-90 and discussed the 
possible nature of this fascinating object. It shows a jet-like
structure with several perfectly aligned knots on both sides of the 
star that seem to be ejected regularly (every 40\,years). This jet 
structure is perpendicular to the disk-like structure seen with HST.
This disk seems to be neutral in hydrogen since strong [O{\sc i}] emission
has been observed that can only come from this disk. During the analysis
of the forbidden emission lines it turned out that C, 
O {\bf and} N need to be depleted to reproduce the observed line luminosities.
This fact is not consistent with any known stellar evolution scenario of single
stars. In addition, a kind of accretion mechanism is certainly necessary to
create the observed knots. From our comparison with other well known 
systems having jets and knots we could exclude the YSO and the symbiotic
nature of Hen 2-90. We cannot exclude Hen 2-90 being an X--ray binary since
only observations with Chandra in the soft X--ray bands have to date been 
performed. But it seems to be clear that some kind of interaction has been
or is still going on in this system.
We want to finish our contribution by raising questions open for further 
discussions and investigations:
\begin{enumerate}
\item No wiggling of the knots around the jet axis -- hint for a close binary?
\item The period of knot ejection of about 40\,years -- if it is due to
binary interaction must it then be an extremely excentric system?
\item The disk like structure seen on the HST image -- an outflowing disk
or the remnant of an earlier common envelope phase?
\item The variety of line profiles -- hint for a complex velocity and
density structure of the circumstellar matter due to ongoing mass transfer, 
accretion and ejection?
\end{enumerate}
Only further observations (e.g. in the hard X--ray bands) will help to
understand and disentangle the real nature of the enigmatic object Hen 2-90.

\begin{acknowledgements}
M.K. acknowledges financial support from the Nederlandse Organisatie voor
Wetenschappelijk Onderzoek grant No.\,614.000.310.
M.B.F. acknowledges financial support from CNPq, Brazil.
\end{acknowledgements}


\begin{thebibliography}{}
\bibitem[Chu et al.(2003)]{c2003} Chu, Y.-H., Gruendl, R. A. \& Guerrero, M. A.
  2003, RMxAC, 15, 62
\bibitem[Costa et al.(1993)]{c1993} Costa, R. D. D., de Freitas Pacheco,
  J. A. \& Maciel, W. J. 1993, A\&A, 276, 184
\bibitem[Henize(1967)]{h1967} Henize, K. G. 1967, ApJS, 14, 125
\bibitem[Kaler \& Jacoby(1991)]{k1991} Kaler, J. B. \& Jacoby, G. H. 1991,
  ApJ, 372, 215
\bibitem[Kraus et al.(2004)]{k2004} Kraus, M., Borges Fernandes, M., 
  de Ara\'{u}jo, F. X. \& Lamers, H. J. G. L. M. 2004, A\&A, {\it submitted}
\bibitem[Kraus \& Lamers(2003)]{k2003} Kraus, M. \& Lamers, H. J. G. L. M., 
  2003, A\&A, 405, 165
\bibitem[Lamers et al.(1998)]{l1998} Lamers, H. J. G. L. M., Zickgraf, F.-J.,
  de Winter, D., Houziaux, L. \& Zorec, J. 1998, A\&A, 340, 117
\bibitem[Livio(1999)]{l1999} Livio, M. 1999, Physics Reports, 311, 225
\bibitem[Sahai \& Nyman(2000)]{sn2000} Sahai, R. \& Nyman, L.-\AA. 2000,
  ApJ, 537, L\,145
\bibitem[Sahai et al.(2002)]{s2002} Sahai, R., Brillant, S., Livio, M. et al. 
  2002, ApJ, 573, L\,123
\bibitem[Schwarz et al.(1992)]{s1992} Schwarz, H. E., Corradi, R. L. M. \& 
  Melnick, J. 1992, A\&AS, 96, 23
\end{thebibliography}
\end{document}